\documentclass[aps,prd,twocolumn,groupedaddress,showpacs]{revtex4} 
\usepackage{bm}
\usepackage[cp866]{inputenc}
\usepackage[english]{babel}

\usepackage{graphicx}

\input epsf

\begin{document}

\title{Nature of the \boldmath $a_0(980)$ Meson in the Light of Photon-Photon
Collisions}
\author{N.~N.~Achasov}
\author{G.~N.~Shestakov}

\affiliation{Laboratory of  Theoretical Physics, S.L. Sobolev
Institute for Mathematics, 630090, Novosibirsk, Russia}


\begin{abstract}
New high-statistics Belle data on the reaction
$\gamma\gamma\to\pi^0\eta$ are analyzed to clarify the two-photon
production mechanisms and the nature of the $a_0(980)$ meson. The
obtained solution for the amplitude $\gamma\gamma\to\pi^0\eta$ is
consistent with the chiral theory expectation for the $\pi\eta$
scattering length, with the strong coupling of the $a_0(980)$ to the
$\pi\eta$, $K\bar K$, and $\pi\eta'$ channels, and with a key role
of the rescattering mechanisms $a_0(980)$\,$\to$\,$(K\bar K+
\pi^0\eta+\pi^0\eta')$\,$\to$\,$\gamma \gamma$ in the
$a_0(980)$\,$\to $\,$\gamma \gamma$ decay. Such a picture argues in
favor of the $q^2\bar q^2$ nature of the $a_0(980)$ meson and is in
agreement with the properties of its partners, the $\sigma_0(600)$
and $f_0(980)$ mesons, in particular, with those that manifest
themselves in $\gamma\gamma\to\pi\pi$. An important role of the
vector meson exchanges in the formation of the nonresonant
background in $\gamma\gamma\to\pi^0\eta$ is also revealed. The
preliminary information on the reaction $\pi^0\eta\to\pi^0\eta$ is
obtained.
\end{abstract}
 \maketitle

\section{INTRODUCTION}

\begin{figure}\centerline{\epsfysize=2.1in 
\epsfbox{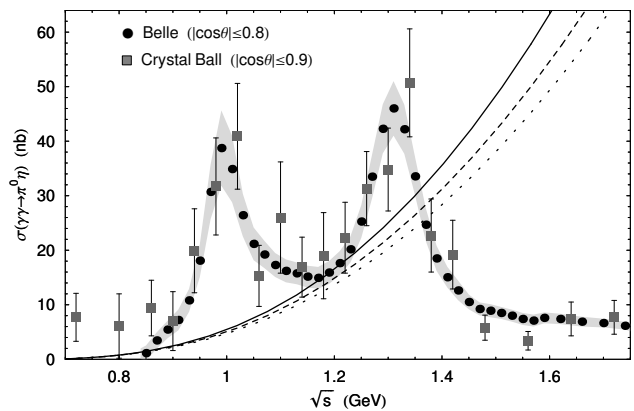}} \vspace{-1mm} \caption{The Belle and
Crystal Ball data for the $\gamma\gamma\to\pi^0\eta$ cross section.
$\theta$ denotes the output angle of the $\pi^0$ (or $\eta$) in the
$\gamma\gamma$ center-of-mass system. The average statistical error
of the Belle data is approximately $\pm0.4$ nb, the shaded band
shows the size of their systematic error. The solid, dashed, and
dotted lines correspond to the total, helicity 0, and $S$ wave
$\gamma\gamma\to\pi^0\eta$ cross sections caused by the elementary
$\rho$ and $\omega$ exchanges for
$|\cos\theta|\leq0.8$.}\label{BelleData}\end{figure}

Recently, the Belle Collaboration obtained new high-statistics data
on the reaction $\gamma\gamma\to\pi^0\eta$ at the KEKB $e^+e^-$
collider \cite{Ue09}. The statistics collected in the Belle
experiment is 3 orders of magnitude higher than in the earlier
experiments performed by the Crystal Ball (336 events) \cite{An86}
and JADE (291 events) \cite{Oe90} Collaborations. The experiments
revealed a specific feature of the $\gamma\gamma\to\pi^0\eta$ cross
section. It turned out sizable in the region between the $a_0(980)$
and $a_2 (1320)$ resonances (see Fig. \ref{BelleData}), which
certainly indicates the presence of additional contributions. These
contributions must be coherent with the resonance ones because the
$\gamma \gamma\to\pi^0\eta$ amplitude in the $\pi^0\eta$ invariant
mass region $\sqrt{s}<1.4$ GeV is dominated by two lowest partial
waves \cite{Ue09}: $S$ and $D_2$ waves
($D_2$\,$\equiv$\,$D_{\lambda=2}$, where $\lambda$ is the absolute
value of the difference between the helicities of the initial
photons). The Belle Collaboration carried out the fit to the
$\gamma\gamma\to\pi^0\eta$ data taking into account interference
between resonance and background contributions \cite{Ue09}. In so
doing the simplified Breit-Wigner functions were used to describe
the resonance contributions with spin $J$\,=\,0 and 2. For example,
in propagators of the $a_0(980)$ and the putative heavy $a_0(Y)$
resonance only the coupling to the $\pi\eta$ channel was taken into
account, and the $a_0(980)$\,$\to$\,$\gamma \gamma$ and
$a_0(Y)$\,$\to$\,$\gamma \gamma$ transition amplitudes were
approximated by constants \cite{Ue09}. The background contributions
were approximated by second order polynomials in $\sqrt{s}$. It
turned out that the description of the $S$ wave requires a smooth
background with the amplitude comparable in magnitude with the
$a_0(980)$ resonance amplitude in its maximum and having the large
imaginary part \cite{Ue09}. As a result the background leads,
practically, to quadrupling the cross section in the $a_0(980)$ peak
region and to filling the dip between the $a_0(980)$ and $a_2
(1320)$ resonances. The origin of such a considerable background in
the $S$ wave is unknown. The imaginary part of the background
amplitude is determined by the contributions of the real
intermediate states $\pi\eta$, $K\bar K$, $\pi\eta'$, and certainly
requires the distinct dynamical decoding.

In this work we shall  show that the experimentally observed pattern
is the result of the combination of many dynamical factors. To
analyze the data, we have significantly developed the model proposed
previously in Ref. \cite{AS88}. The basis for this model is an idea
of what the $a_0(980)$ resonance can be as a suitable candidate in
four-quark states \cite{Ja77, ADS80}. There exists a set of
important evidence in favor of the four-quark nature of the
$a_0(980)$; see, for example, Refs.
\cite{ADS80,ADS82,AI89,Ac98,AK03,Ac08}. As a result of the performed
analysis, we have elucidated that the Belle data are in close
agreement with the case of strong coupling the $a_0(980)$ to the
$\pi\eta$, $K\bar K$, and $\pi\eta'$ channels, and with a key role
of the rescattering mechanisms $a_0(980)$\,$\to$\,$(K\bar K+
\pi^0\eta+\pi^0\eta')$\,$\to$\,$\gamma \gamma$, i.e., four-quark
transitions, in the decay $a_0(980)$\,$\to$\,$\gamma\gamma$
\cite{FN1,AS05,AS08,AS09ar,Mo07,Ue08}. Furthermore, the nontrivial
evidence has been found for the important role of the nonresonant
production $\gamma\gamma $\,$\to$\,$\pi^0\eta$ via vector meson
exchanges. The model information on the reaction
$\pi^0\eta\to\pi^0\eta$, which we have extracted from the Belle
data, is in reasonable agreement with the expectations based on
chiral dynamics \cite{Os70,BKM91,BFS}.

\section{DYNAMICAL MODEL FOR \boldmath $\gamma\gamma$\,$\to$\,$\pi^0\eta$}

To analyze the data, we use a model for the helicity, $M_\lambda$,
and corresponding partial, $M_{\lambda J}$, amplitudes of the
reaction $\gamma\gamma$\,$\to$\,$ \pi^0\eta$, where the
electromagnetic Born contributions from $\rho$, $\omega$, $K^*$, and
$K$ exchanges modified by form factors and strong elastic and
inelastic final-state interactions in $\pi^0\eta$, $\pi^0\eta'$,
$K^+K^-$, and $K^0\bar K^0$ channels, as well as the contributions
due to the direct interaction of the resonances with photons, are
taken into account:
\begin{figure} \centerline{\epsfxsize=3in 
\epsfbox{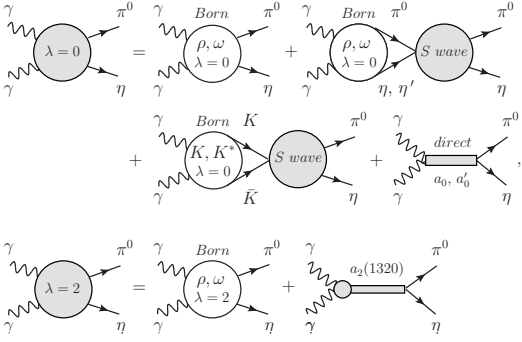}} 
\caption {Diagrammatical representation for the helicity amplitudes
$\gamma\gamma$\,$\to$\,$\pi^0\eta$.}\label{Amplitudes}\end{figure}
\begin{figure} \begin{center}\begin{tabular}{c}{\epsfxsize=2.7in 
\epsfbox{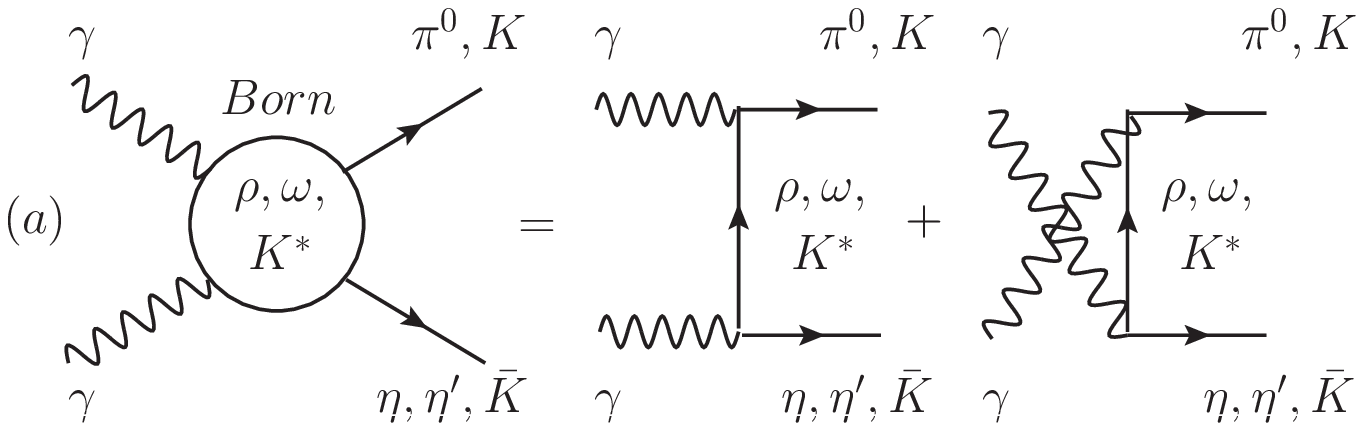}}\\ {\epsfxsize=2.7in 
\epsfbox{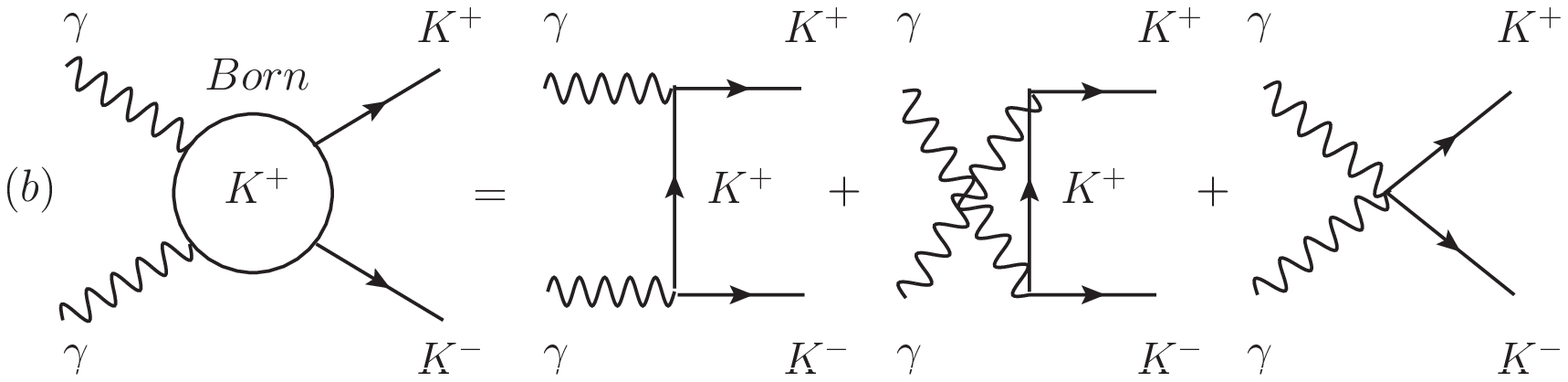}}\end{tabular}\end{center} \vspace{-2mm}
\caption {The Born $\rho$, $\omega$, $K^*$, $K$ exchange diagrams
for $\gamma\gamma$\,$\to$\,$\pi^0\eta$, $\gamma\gamma$\,$
\to$\,$\pi^0\eta'$, and $\gamma\gamma$\,$\to$\,$K\bar K$.
}\label{BornVAKK}\end{figure}
\begin{eqnarray}
& M_0(\gamma\gamma\to\pi^0\eta;s,\theta)=M^{\mbox{\scriptsize{Born}}
\,V}_0(\gamma\gamma\to\pi^0\eta;s,\theta) & \nonumber\\
& +\widetilde{I}^V_ {\pi^0\eta}(s)\,T_{\pi^0\eta\to\pi^0\eta}(s)+
\widetilde{I}^V_ {\pi^0\eta'}(s)\,T_{\pi^0\eta'\to\pi^0\eta}(s) &
\nonumber\\ &
+\left(\widetilde{I}^{K^{*+}}_{K^+K^-}(s)-\widetilde{I}^{K^{*0}}_{K^0\bar
K^0}(s)+\widetilde{I}^{K^+}_{K^+K^-}(s)\right) & \nonumber\\ &
\times\,T_{K^+K^-\to \pi^0\eta}(s)+M^{\mbox
{\scriptsize{direct}}}_{\mbox{\scriptsize{res}}}(s), &
\label{M0} \\[0.14cm]
& M_2(\gamma\gamma\to\pi^0\eta;s,\theta)=M^{\mbox{
\scriptsize{Born}}\,V}_2(\gamma\gamma\to\pi^0\eta;s,\theta)&
\nonumber\\ & +80\pi d^2_{20}(\theta) M_{\gamma\gamma\to
a_2(1320)\to\pi^0\eta}(s), & \label{M2}
\end{eqnarray}
$d^2_{20}(\theta)$\,=\,$(\sqrt{6}/4)\sin^2 \theta$; the diagrams
corresponding to these amplitudes are shown in Figs.
\ref{Amplitudes} and \ref{BornVAKK}.

The first terms in the right-hand parts of Eqs. (\ref{M0}) and
(\ref{M2}) represent the real Born helicity amplitudes caused by the
contributions of the $\rho$ and $\omega$ exchange mechanisms. These
contributions are equal in magnitude and in sign. With this note we
write \cite{AS88,AS92}
\begin{eqnarray}\label{MBornV0} && \mbox{\qquad\ \ \ }
M^{\mbox{\scriptsize{Born}}
\,V}_0(\gamma\gamma\to\pi^0\eta;s,\theta)\nonumber \\
&& =2g_{\omega\pi\gamma}g_{\omega\eta\gamma}\frac{s}{4}\left[
\frac{tG_\omega(s,t)}{t-m^2_\omega}+\frac{uG_\omega(s,
u)}{u-m^2_\omega}\right],\mbox{\ } \\[0.14cm]
\label{MBornV2} && \mbox{\qquad\ \ \ }M^{\mbox{\scriptsize{Born}}
\,V}_2(\gamma\gamma\to\pi^0\eta;s,\theta)\nonumber \\
&& =2g_{\omega\pi\gamma}g_{\omega\eta\gamma}\frac{m^2_\pi
m^2_\eta-tu}{4}\left[
\frac{G_\omega(s,t)}{t-m^2_\omega}+\frac{G_\omega(s,
u)}{u-m^2_\omega}\right],\mbox{\ } \end{eqnarray} where $t$ and $u$
are the Mandelstam variables for the reaction
$\gamma\gamma\to\pi^0\eta$,
$g_{\omega\eta\gamma}$\,=\,$\frac{1}{3}g_{\omega\pi\gamma}
\sin(\theta_i-\theta_P)$,
$g^2_{\omega\pi\gamma}$\,=\,$12\pi\Gamma_{\omega\to\pi\gamma}
[(m^2_\omega-m^2_\pi)/(2m_\omega)]^{-3}\approx0.519$ GeV$^{-2}$
\cite{PDG08}, the ``ideal'' mixing angle $\theta_i$\,=\,35.3$^\circ
$, the pseudoscalar mixing angle $\theta_P$ is a free parameter,
$G_\omega(s,t)$ and $G_\omega(s, u)$ are the form factors in the $t$
and $u$ channels [for the elementary $\rho$ and $\omega$ exchanges
$G_\omega(s,t)$\,=\,$G_\omega(s, u)$\,=\,1]. In the corresponding
Born amplitudes for $\gamma\gamma\to\pi^0\eta'$
$g_{\omega\eta'\gamma}
$\,=\,$\frac{1}{3}g_{\omega\pi\gamma}\cos(\theta_i-\theta_P)$. The
amplitudes $M^{\mbox{\scriptsize{Born}}\,K^*}_\lambda(\gamma\gamma
\to K\bar K;s,\theta)$ for the $K^*$ exchanges result from Eqs.
(\ref{MBornV0}), (\ref{MBornV2}) with the use of substitutions
$m_\omega$\,$\to$\,$m_{K^*}$, $G_\omega$\,$\to$\,$G_{K^*}$,
$m_\pi$\,$\to$\,$m_K$, $m_\eta$\,$\to$\,$m_K$, and
$2g_{\omega\pi\gamma}g_{\omega\eta\gamma}$\,$\to$\,$g^2_{K^*K\gamma}$,
where $g^2_{K^{*+}K^+\gamma}\approx0.064$ GeV$^{-2}$ and
$g^2_{K^{*0} K^0\gamma} \approx0.151$ GeV$^{-2}$ \cite{PDG08}.

Note that the Born sources of the reaction
$\gamma\gamma\to\pi^0\eta$ corresponding to the $t$ and $u$ channel
exchanges with the quantum numbers of the vector $\rho$ and $\omega$
mesons [as well as with those of the axial-vector $b_1(1235)$ and
$h_1(1170)$ mesons] are poorly understood in the nonasymptotic
energy region of interest. One can say only that the elementary
$\rho$ and $\omega$ exchanges, whose contributions to the
$\gamma\gamma\to\pi^0\eta$ cross section (primarily to the $S$ wave)
increase very rapidly with the energy, are not observed in the
experiments; see Fig. \ref{BelleData}. This fact was explained in
Ref. \cite{AS92}. The appropriate Reggeization of the elementary
exchanges with higher spins strongly reduces dangerous
contributions. Such a reduction has to take place already in the
region of 1--1.5 GeV. Therefore, in applications to the real case,
it is natural to use the form factors of the Regge type:
$G_\omega(s,t)$\,=\,$\exp[(t-m^2_\omega)b_\omega (s)]$, $G_\omega(s,
u)$\,=\,$\exp[(u-m^2_\omega)b_\omega(s)]$, where we put
$b_\omega(s)$\,=\,$b^0_\omega+(\alpha'_\omega/4) \ln[1+(s/s_0)^4]$,
$b^0_\omega$\,=\,0, $\alpha'_\omega$\,=\,0.8 GeV$^{-2}$ and
$s_0$\,=\,1 GeV$^2$ (form factors for the $K^*$ exchange result from
the above by substitution of $K^*$ for $\omega$).

As for the $b_1(1235)$ and $h_1(1170)$ exchanges, their amplitudes
have the form analogous to expressions (\ref{MBornV0}) and
(\ref{MBornV2}) except for the common sign in the amplitude with
helicity 0. The available (rather poor) information on the coupling
constant gives $g^2_{h_1\pi\gamma}\approx9g^2_{b_1\pi\gamma}\approx
$\,0.34\,GeV$ ^{-2}$ \cite{PDG08}. The coupling constant, higher
mass, and more damping form factor lead to the fivefold (at least)
suppression of the axial-vector exchanges in comparison with the
vector ones, which is why we neglect their contributions.

The terms in Eq. (\ref{M0}), proportional to the $S$ wave amplitudes
of hadronic reactions, $T_{\pi^0\eta\to\pi^0\eta}(s)$,
$T_{\pi^0\eta'\to\pi^0\eta}(s)$, and $T_{K^+K^-\to\pi^0\eta}(s)$,
are due to the rescattering mechanisms. In amplitudes $T$ we take
into account the contribution from mixed $a_0(980)$ and (heavy)
$a_0(Y)$ resonances (hereinafter referred to as $a_0$ and $a'_0$,
respectively), and the background contributions. The amplitudes $T$
are given by
\begin{eqnarray} && T_{\pi^0\eta\to\pi^0\eta}(s)=T_0^1 (s)=
\frac{\eta^1_0(s)e^{2i \delta^1_0(s)}-1}{2i\rho_{\pi\eta}(s)}
\nonumber \\ && \mbox{\qquad}
=T_{\pi\eta}^{bg}(s)+e^{2i\delta_{\pi\eta}^{bg}
(s)}T^{res}_{\pi^0\eta\to\pi^0\eta}(s)\,, \label{Tpietapieta}\\
&& T_{\pi^0\eta'\to\pi^0\eta}(s)=T^{res}_{\pi^0\eta'
\to\pi^0\eta}(s)\,e^{i[\delta_{\pi\eta'}^{bg}(s)+\delta_{\pi\eta}^{bg}
(s)]},\label{Tpieta1pieta} \\ && T_{K^+K^-\to\pi^0\eta}(s)=
T^{res}_{K^+K^-\to\pi^0 \eta}(s)\,e^{i[\delta_{K\bar
K}^{bg}(s)+\delta_{\pi\eta}^{bg}(s)]},\ \ \ \ \label{TKKpieta}
\end{eqnarray} where
$T_{\pi\eta}^{bg}(s)=(e^{2i\delta_{\pi\eta}^{bg}(s)}-1)/(2i
\rho_{\pi\eta}(s))$, $T^{res}_{\pi^0\eta\to\pi^0\eta}(s)
=(\eta^1_0(s)e^{2i\delta_{\pi\eta}^{res}(s)}-1)/(2i\rho_{
\pi\eta}(s))$, $\delta^1_0(s)=\delta_{\pi\eta}^{bg}(s)+
\delta_{\pi\eta}^{res}(s)$, $\rho_{ab}(s)$\,=\,$
\sqrt{(1-m^{(-)2}_{ab}/s)(1-m^{(+)2}_{ab}/s)}$,
$m_{ab}^{(\pm)}$\,=\,$m_b\pm m_a$ ($ab$\,=\,$\pi\eta$, $K^+K^-$,
$K^0\bar K^0 $, $\pi\eta'$), $\delta_{\pi\eta}^{bg}(s)$,
$\delta_{\pi\eta'}^{bg}(s)$, and $\delta_{K\bar K}^{bg}(s)$ are the
phase shifts of the elastic background contributions in the channels
$\pi\eta$, $\pi\eta'$, and $K\bar K$ with isospin $I=1$,
respectively.

The amplitudes of the $a_0$\,--\,$a'_0$ resonance complex in Eqs.
(\ref{Tpietapieta}), (\ref{Tpieta1pieta}), and (\ref{TKKpieta}) have
the form \cite{ADS80,ADS79,AK06,AS08}:
\begin{equation}\label{Tres}
T^{res}_{ab\to\pi^0\eta}(s)=\frac{g_{a_0ab}\Delta_{a'_0}(s)+g_{a'_0ab
}\Delta_{a_0}(s)}{16\pi[D_{a_0}(s)D_{a'_0}(s)-\Pi^2_{a_0a'_0}(s)]}\,,
\end{equation} where
$\Delta_{a'_0}(s)$\,=\,$D_{a'_0}(s)g_{a_0\pi^0\eta}+\Pi_{a_0a'_0}
(s)g_{a'_0\pi^0\eta}$ and $\Delta_{a_0}(s)$
=\,$D_{a_0}(s)g_{a'_0\pi^0\eta}+\Pi_{a_0a'_0} (s)g_{a_0\pi^0\eta}$,
$g_{a_0ab}$ and $g_{a'_0ab}$ are the coupling constants, the
propagator of the $a_0$ (and similarly $a'_0$) resonance is
\begin{equation}\label{Prop-a0}\frac{1}{D_{a_0}(s)}=\frac{1}{m^2_{a_0}-s+
\sum_{ab}[\mbox{Re}\Pi^{ab}_{a_0}(m^2_{a_0})-\Pi^{ab}_{a_0}(s)]},\end{equation}
where $\mbox{Re}\Pi^{ab}_{a_0}(s)$ is determined by a singly
subtracted dispersion integral of
$\mbox{Im}\Pi^{ab}_{a_0}(s)=\sqrt{s}\Gamma_{a_0\to
ab}(s)=g^2_{a_0ab}\rho_{ab}(s)/(16\pi)$;
$\Pi_{a_0a'_0}(s)$\,=\,$C+\sum_{ab}(g_{a'_0ab}/g_{a_0ab})\Pi^{ab}_{a_0}(s)$,
where $C$ is the mixing parameter. The explicit form of the
polarization operators $\Pi^{ab}_{a_0}(s)$ has been written out in
Refs. \cite{ADS80,AK03,AK06,ADS80a,AS09}.

The amplitude
\begin{equation}\label{Mdirect} M^{\mbox{\scriptsize{direct}}}_
{\mbox{\scriptsize{res}}}(s)=
s\frac{g^{(0)}_{a_0\gamma\gamma}\Delta_{a'_0}(s)+
g^{(0)}_{a'_0\gamma\gamma}\Delta_{a_0}(s)}
{D_{a_0}(s)D_{a'_0}(s)-\Pi^2_{a_0a'_0}(s)}e^{i\delta_{\pi\eta}^{bg}(s)}
\end{equation} in Eq. (\ref{M0})
describes the $\gamma\gamma\to\pi^0\eta$ transition caused by the
direct coupling constants of the $a_0$ and $a'_0$ resonances to
photons $g^{(0)}_{a_0\gamma\gamma}$ and $g^{(0)}_{a'_0\gamma\gamma
}$; the factor $s$ appears due to the gauge invariance.

Equation (\ref{M0}) implies that the amplitudes
$T_{\pi^0\eta\to\pi^0\eta}(s)$, $T_{\pi^0\eta'\to\pi^0\eta}(s)$, and
$T_{K\bar K\to\pi^0\eta}(s)$ in $\gamma\gamma\to\pi^0\eta\to\pi^0
\eta$, $\gamma\gamma\to\pi^0\eta'\to\pi^0\eta$, and $\gamma\gamma\to
K\bar K\to\pi^0\eta$ rescattering loops (see Fig. \ref{Amplitudes})
lie on the mass shell \cite{FNN1}. In so doing, the functions
$\widetilde{I}^V_{\pi^0\eta}(s)$, $\widetilde{I}^V_{\pi^0\eta'}(s)$,
$\widetilde{I}^{K^*}_{K\bar K}(s)$, and
$\widetilde{I}^{K^+}_{K^+K^-}(s)$ are the amplitudes of the triangle
loop diagrams describing the transitions
$\gamma\gamma$\,$\to$\,$ab$\,$\to$\,({\it scalar state with a mass
}\,=\,$\sqrt{s}$), in which the meson pairs $\pi^0\eta$,
$\pi^0\eta'$, and $K\bar K$ are produced by the electromagnetic Born
sources (see Fig. \ref{BornVAKK}). For example, the function
\begin{eqnarray}\label{IKK} \widetilde{I}^{K^+}_{K^+K^-}(s)=\frac{s}
{\pi}\mbox{\qquad\qquad\qquad\ }\nonumber \\
\times\int\limits^\infty_{4m^2_{K^+}}\frac{\rho_{K^+K^-}(s')M^{
\mbox{\scriptsize{Born}}\,K^+}_{00}(\gamma\gamma\to
K^+K^-;s')}{s'(s'-s-i\varepsilon)}ds',&& \end{eqnarray} where
$M^{\mbox {\scriptsize{Born}}\,K^+}_{00}(\gamma\gamma
$\,$\to$\,$K^+K^-;s')$ is the $S$ wave of the Born charged one-kaon
exchange amplitude (see Refs. \cite{FN2,Po86,Jo86,MPBB,AS94} for
details). The functions $\widetilde{I}^V_{\pi^0\eta}(s)$,
$\widetilde{I}^V_{\pi^0\eta'}(s)$, and $\widetilde{I}^{K^*}_{K\bar
K}(s)$ are calculated in a similar way. The amplitude
$M_0(\gamma\gamma\to\pi^0\eta;s,\theta)$ constructed in such a way
satisfies the Watson theorem in the elastic region \cite{AS05,AS08}.

For the $a_2(1320)$ production amplitude [see Eq. (\ref{M2})], we
use the following simple parametrization:
\begin{eqnarray}\label{Ma2}
M_{\gamma\gamma\to a_2(1320)\to\pi^0\eta}(s)\mbox{\qquad\qquad}
\nonumber \\ =\frac{G_{a_2}(s)
\sqrt{s\Gamma^{\mbox{\scriptsize{tot}}}_{a_2}(s)B(a_2\to\pi\eta)
/\rho_{\pi\eta}(s)}}{m^2_{a_2}-s-i
\sqrt{s}\Gamma^{\mbox{\scriptsize{tot}}}_{a_2}(s)}\,. &&
\end{eqnarray} Here
\begin{eqnarray} G_{a_2}(s)=
\sqrt{\Gamma^{(0)}_{a_2\to\gamma\gamma}(s)}+i\frac{M_{22}^{
\mbox{\scriptsize{Born}}\,V}(\gamma\gamma\to\pi^0\eta;s)}{16\pi}\ &&
\nonumber \\ \times\sqrt{\rho_{\pi\eta}(s)
\Gamma^{\mbox{\scriptsize{tot}}}_{a_2}(s)
B(a_2\to\pi\eta)}\ ,\mbox{\qquad\qquad} \label{Ga2} \\[0.1cm]
\Gamma^{\mbox{\scriptsize{tot}}}_{
a_2}(s)=\Gamma^{\mbox{\scriptsize{tot}}}_{a_2}\frac{m^2_{a_2}}{s}
\frac{q^5_{\pi\eta}(s)}{q^5_{\pi\eta}(m^2_{a_2})}\frac{D_2(q_{\pi
\eta}(m^2_{a_2})r_{a_2})}{D_2(q_{\pi \eta}(s)r_{a_2}) }\,,\ \  &&
\label{Ga2tot}\end{eqnarray} were
$q_{\pi\eta}(s)$\,=\,$\sqrt{s}\rho_{ \pi\eta}(s)/2$,
$D_2(x)$\,=\,$9+3x^2+x^4$ and $r_{a_2}$ is the interaction radius.
By definition $\Gamma_{a_2\to\gamma\gamma}(s)$\,=\,$|G_{a_2}(s)|^2$
and $\Gamma^{(0)}_{a_2\to\gamma\gamma}(s)=(\frac{\sqrt{s}
}{m_{a_2}})^3 \Gamma^{(0)}_{a_2\to\gamma\gamma}$. The second term in
$G_{a_2}(s)$ corresponds to the
$a_2(1270)$\,$\to$\,$\pi^0\eta$\,$\to$\,$\gamma\gamma$ transition
with the real $\pi^0$ and $\eta$ mesons in the intermediate state.
The estimates show that this term is less than 3\% in the amplitude
and can be omitted. Recall that the observed two-photon decays
widths of the $f_2(1270)$ and $a_2(1320)$ mesons \cite{PDG08}
satisfy rather well the relation $\Gamma_{f_2\to\gamma\gamma}/
\Gamma_{a_2\to \gamma\gamma }$\,=\,$25/9$, which holds in the naive
$q\bar q$ model for the direct annihilation transitions $q\bar
q$\,$\to$\,$\gamma \gamma$.

The normalization of the amplitudes $M_\lambda$ is specified by
their relation with the cross section $\sigma=\sigma_0+\sigma_2$,
where $\sigma_\lambda= \frac{\rho_{\pi\eta}(s)}{64\pi
s}\int|M_\lambda|^2d\cos\theta$.

\section{RESULTS OF THE DATA DESCRIPTION}

\begin{figure}\centerline{\epsfxsize=3.7in 
\epsfbox{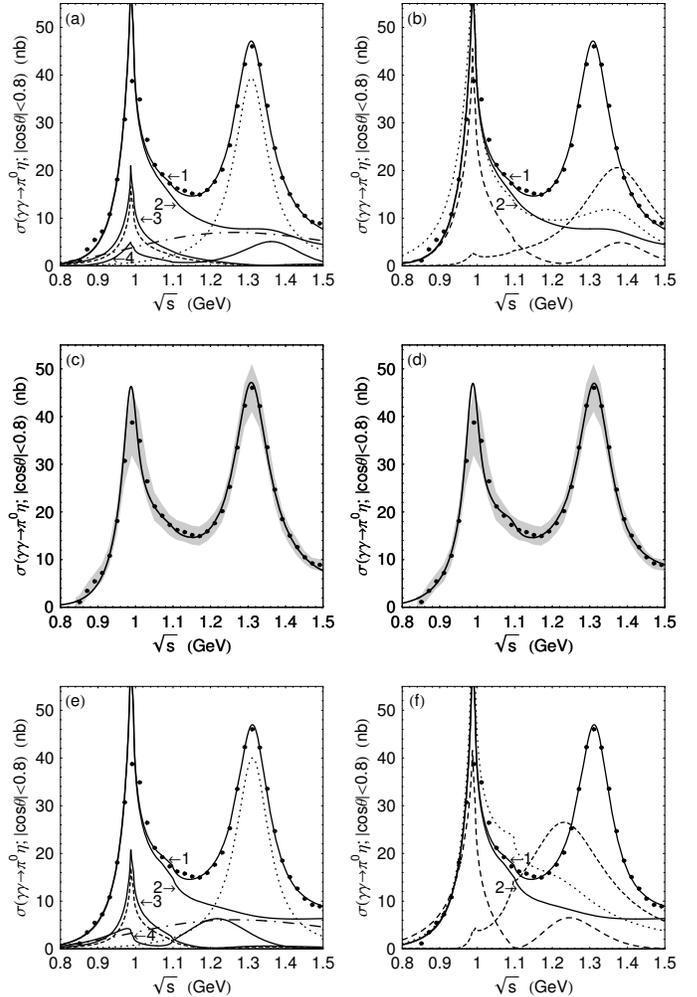}}\vspace{-4mm} \caption{Fits to the Belle
data. The curves in (a), (b), and (c) correspond to variant 1. (a)
The total cross section $\gamma\gamma$\,$\to$\,$\pi^0\eta$ (solid
line 1), its helicity 0 (solid line 2), and helicity 2 (dotted line)
parts, the contribution from the
$\gamma\gamma$\,$\to$\,$K^+K^-$\,$\to$\,$\pi^0\eta$ rescattering
with intermediate $K^+K^-$ produced via the Born $K$ exchange (solid
line 3), the contribution from the $\gamma\gamma$\,$\to$\,$K\bar
K$\,$\to$\,$\pi^0\eta$ rescattering with intermediate $K\bar K$
produced via the Born $K$ and $K^*$ exchanges (dashed line), the
contribution from the Born $\rho$ and $\omega$ exchange amplitude
with $\lambda$\,=0 (dot-dashed line), and the combined contribution
from the Born one and the $S$ wave
$\gamma\gamma$\,$\to$\,$(\pi^0\eta+\pi^0\eta')$\,$ \to$\,$\pi^0\eta$
rescattering (solid line 4); see Figs. (\ref{Amplitudes}) and
(\ref{BornVAKK}). (b) The solid lines 1 and 2 are the same as in
(a), the short-dashed line corresponds to the contribution from the
$\gamma\gamma$\,$\to$\,$\pi^0\eta$ direct transition amplitude for
the $a_0-a'_0$ resonance complex [see Eq. (\ref{Mdirect})], the
dotted line shows the total contribution from the $a_0-a'_0$
resonance one, and the long-dashed line corresponds to the helicity
0 cross section without the contribution from the direct transition
amplitude. The resulting solid line in (c) corresponds to the solid
line 1 in (a) [or (b)], folded, in the region
0.84\,GeV\,$<$\,$\sqrt{s} $\,$<$\,1.15\,GeV, with a Gaussian with
$\sigma$\,=\,10\,MeV mass resolution; the shaded band shows the size
of the systematic error of the data. (d), (e), and (f) The same as
(c), (a), and (b), respectively, for variant 2.}\label{ABCDEF}
\end{figure}

Figures \ref{ABCDEF}(a)--\ref{ABCDEF}(c) illustrate the results of
one of the fitting variants to the Belle data on the cross section
$\gamma\gamma\to\pi^0\eta$ (hereafter variant 1); the corresponding
parameter values are collected in Table \ref{Tab1}. As is seen from
Fig. \ref{ABCDEF}(c), the resulting fitted curve agrees quite
satisfactorily with the data within their systematic errors. Such an
agreement allows definite conclusions concerning the main dynamical
constituents of the $\gamma\gamma$\,$ \to$\,$\pi^0\eta$ reaction
mechanism. The contributions of these constituents are shown in
detail in Figs. \ref{ABCDEF}(a) and \ref{ABCDEF}(b).

We begin with the contribution of the inelastic rescattering
$\gamma\gamma$\,$\to$\,$K^+K^-$\,$\to$\,$\pi^0\eta$, where the
intermediate $K^+K^-$ pair is produced via the charged one-kaon
exchange mechanism [see Fig. \ref{BornVAKK}(b)]. This mechanism
predicts the natural scale for the $a_0(980)$ production cross
section in $\gamma\gamma$\,$\to$\,$\pi^0\eta$ \cite{AS88,AS92,AS09,
FN3}. The maximum of the cross section $\gamma\gamma$\,$
\to$\,$K^+K^-$\,$\to$\,$a_0(980) $\,$\to$\,$\pi^0\eta$ is basically
controlled by the product of the parameter $R_{a_0}$\,=\,$
g^2_{a_0K^+K^-} /g^2_{a_0 \pi\eta}$ and the value $|\widetilde{I}
^{K^+}_{K^+K^-}(4m^2_{K^+})|^2$. Its estimate gives
$\sigma(\gamma\gamma$\,$\to$\,$K^+K^-$\,$\to$\,$ a_0(980)$\,$
\to$\,$\pi^0\eta;|\cos\theta|\leq0.8)\approx0.8\times1.4\alpha^2
R_{a_0}/m^2_{a_0}\approx24$\,nb\,$\times R_{a_0}$ (here we neglect
the heavy $a'_0 $ resonance contribution). The specific feature of
this mechanism is the strong energy dependence of the
$a_0(980)$\,$\to$\,$K^+K^-$\,$ \to$\,$\gamma\gamma$ decay width,
which is conditioned by the function $\widetilde{I}^{K^+}_{K^+K^-}
(s)$ \cite{AS88,AS05,AS08,AS09,FN2}. $\widetilde{I}^{K^+}_{K^+K^
-}(s)$ decreases sharply just below the $K^+K^-$ threshold that
leads to the noticeable additional narrowing of the $a_0(980)$ peak
in the $\gamma\gamma\to\pi^0\eta$ channel; namely, its effective
width turns out to be about 40 MeV at $\Gamma_{a_0\to\pi\eta}(m^2_{
a_0})\approx $\,200 MeV (see Refs. \cite{ADS80,AS88} for details).
The $\gamma\gamma $\,$\to$\,$K^+K^-$\,$\to$\,$\pi^0\eta$
rescattering contribution to the cross section $\gamma\gamma
\to\pi^0\eta$ is shown by solid line 3 in Fig. \ref{ABCDEF}(a). The
$K^*$ exchange narrows slightly the $a_0(980)$ peak [see the dashed
line under solid line 3 in Fig. \ref{ABCDEF}(a)].

It is clear that the $\gamma\gamma $\,$\to$\,$ K\bar
K$\,$\to$\,$\pi^0\eta$ rescattering mechanism alone cannot describe
the data in the $a_0(980)$ resonance region. The observed cross
section can be obtained by adding the Born $\rho$ and $\omega$
exchange contribution, modified by the $S$ wave rescattering $\gamma
\gamma$\,$\to$\,$(\pi^0\eta+\pi^0\eta')$\,$ \to$\,$\pi^0\eta$, and
the amplitude of Eq. (\ref{Mdirect}), caused by the direct
transitions of the $a_0$ and $a'_0$ resonances into photons. Each of
the contributions of these two mechanisms are not too large in the
$a_0(980)$ region (see solid line 4 in Fig. \ref{ABCDEF}(a) for the
first of them and the short-dashed line in Fig. \ref{ABCDEF}(b) for
the second). But the main thing is that their coherent sum with the
contribution of the inelastic rescattering $\gamma\gamma$\,$\to$\,$
K\bar K$\,$\to$\,$\pi^0\eta$ (see the diagrams for the amplitude
with $\lambda$\,=\,0 in Fig. \ref{Amplitudes}) leads to the
considerable enhancement of the $a_0(980)$ resonance manifestation
[see solid line 2 in Fig. \ref{ABCDEF}(a)]. Recall that all the $S$
wave contributions to the amplitude $\gamma\gamma$\,$\to$\,$\pi^0
\eta$ have the equal phase for $\sqrt{s}$ below the $K^+K^-$
threshold, in accord with the Watson theorem.

\begin{figure}\centerline{\epsfysize=3.5in 
\epsfbox{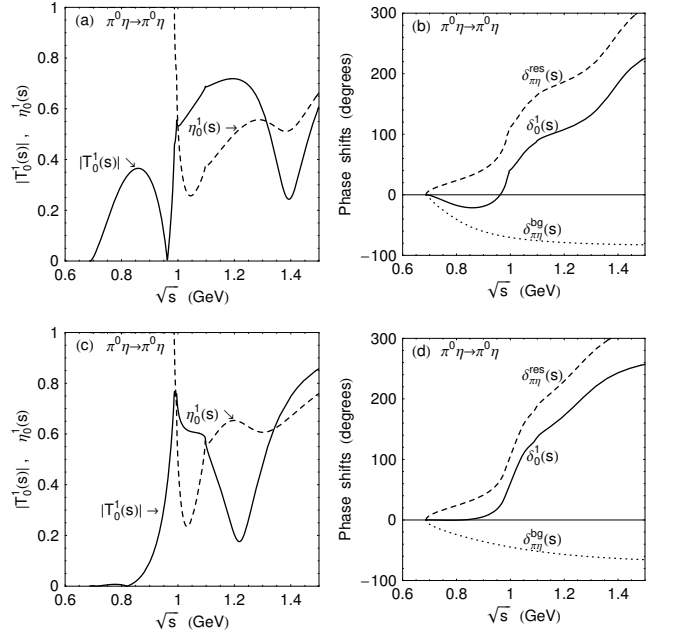}} \vspace{-2mm} \caption{Modulus of $T^1_0(s)$,
inelasticity $\eta^1_0(s)$ (a), and phase shifts (b) of the $S$ wave
amplitude $\pi^0\eta\to\pi^0\eta$ for variant 1 (here
$a^1_0$\,=\,0.0098); (c) and (d) the same for variant 2 (here
$a^1_0$\,=\,0.0066).}\label{TEPS}
\end{figure}

The important role played by the background elastic amplitude of
$\pi^0\eta$ scattering, $T^{bg}_{\pi\eta}(s)$, should be noted; see
Eq. (\ref{Tpietapieta}). First, the choice of the negative
background phase shift $\delta^{bg}_{\pi\eta}(s)$ in
$T^{bg}_{\pi\eta}(s)$ [see Fig. \ref{TEPS}(b)] allows the agreement
of the $\pi\eta$ scattering length in the considered model with the
estimates based on current algebra \cite{Os70} and chiral
perturbation theory \cite{BKM91}, according to which $a^1_0$\,(in
units of $m^{-1}_\pi$)\,$\approx$\,$0.005-0.01$. There takes place a
compensation in $a^1_0$ between the resonance contribution (about
0.3) and the background one \cite{FN4,AS94a}. Second, the
considerable negative value of $\delta^{bg}_{\pi\eta}(s)$ near 1 GeV
provides the resonance like behavior of the cross section shown by
solid line 4 in Fig. \ref{ABCDEF}(a). The characteristics for the
$S$ wave amplitude $\pi^0\eta$\,$\to$\,$ \pi^0\eta$ corresponding to
variant 1 are represented in Figs. \ref{TEPS}(a) and \ref{TEPS}(b).

\begin{table}
  \centering
  \caption{Fitted parameters. The systematic errors of the data are taken into
  account.}\label{Tab1}
\begin{tabular}{|l|l|l|}
  \hline
  $\mbox{\qquad}\,$Variant & $\mbox{\qquad\ \ }$ 1 & $\mbox{\qquad\ \ }$ 2 \\
  \hline
  $\ m_{a_0}            $ (GeV)    & \ 0.9845$^{+0.001}_{-0.004}$   & \ 0.9855$^{+0.0015}_{-0.002}$ \\
  $\ g_{a_0\pi\eta}     $ (GeV)    & \ 4.23$^{+0.25}_{-0.08}$     & \ 3.99$^{+0.17}_{-0.08}$   \\
  $\ g_{a_0K^+K^-}      $ (GeV)    & \ 3.79$^{+0.14}_{-0.71}$     & \ 3.58$\pm0.28$   \\
  $\ g_{a_0\pi\eta'}    $ (GeV)    & \ $-2.13^{+0.4}_{-0.28}$    & \ $-2.34^{+0.32}_{-0.25}$ \\
  $\ g^{(0)}_{a_0\gamma\gamma}$ (GeV$^{-1}$)  & \ $(1.77^{+0.29}_{-0.11})\times10^{-3}$
  & \ $(1.77^{+0.13}_{-0.11})\times10^{-3}$  \\
  \hline
  $\ m_{a'_0}            $ (GeV)    & \ 1.4$^{+0.02}_{-0.01}$    & \ 1.276$\pm0.01$ \\
  $\ g_{a'_0\pi\eta}     $ (GeV)    & \ 3.3$\pm 0.16$            & \ 3.55$^{+0.14}_{-0.1}$  \\
  $\ g_{a'_0K^+K^-}      $ (GeV)    & \ 0.28$\pm 0.3$           & \ 0.48$\pm0.32$  \\
  $\ g_{a'_0\pi\eta'}    $ (GeV)    & \ 2.91$^{+0.35}_{-0.49}$   & \ 2.91$^{+0.29}_{-0.32}$  \\
  $\ g^{(0)}_{a'_0\gamma\gamma}$ (GeV$^{-1}$)  & \ $(-11.5^{+0.3}_{-0.4})\times10^{-3}$
  & \ $(-11.5^{+0.3}_{-0.2})\times10^{-3}$  \\
  \hline
  $\ C              $ (GeV$^2$)       & \ 0.06$^{+0.025}_{-0.03}$     & \ 0.041$\pm0.01$     \\
  \hline
  $\ c_0            $                 & \ $-0.6\pm0.03$ & \ $-0.63\pm0.03$  \\
  $\ c_1            $ (GeV$^{-2}$)    & \ $-6.48^{+1.5}_{-0.65}$  & \ $-1.46^{+0.15}_{-0.32}$   \\
  $\ c_2            $ (GeV$^{-4}$)    & \ 0.12$^{+0.3}_{-0.12}$    & \ 0.095$^{+0.30}_{-0.095}$     \\
  $\ f_{K\bar K}    $ (GeV$^{-1}$)    & \ $-0.37^{+0.13}_{-0.3}$  & \ $-0.37^{+0.13}_{-0.25}$   \\
  $\ f_{\pi\eta'}   $ (GeV$^{-1}$)    & \ 0.28$^{+0.45}_{-0.3}$     & \ 0.3$^{+0.4}_{-0.6}$       \\
  \hline
  $\ \theta_{P}     $ (degree)        & \ $-24^{+3}_{-5}$ & \ $-18^{+3}_{-6}$\\
  \hline
  $\ m_{a_2}        $ (GeV)  & \ 1.322$^{+0.005}_{-0.003}$  & \ 1.323$^{+0.004}_{-0.003}$  \\
  $\ \Gamma^{\mbox{\scriptsize{tot}}}_{a_2}$ (GeV) & \ 0.116$^{+0.011}_{-0.007}$ & \ 0.111$^{+0.007}_{-0.009}$ \\
  $\ \Gamma^{(0)}_{a_2\to\gamma\gamma}$ (keV) & \ 1.053$\pm 0.048$ & \ 1.047$\pm0.005$ \\
  $\ r_{a_2}        $ (GeV$^{-1}$)  & \ 1.9$^{+0.9}_{-1.9}$ & \ 3$^{+2}_{-0.7}$\\
  \hline
  $\ \chi^2/ndf     $                 & \ $5.1/13=0.39$ & \ $4.8/13=0.37$\\
  \hline
\end{tabular}
\end{table}

For the background phase shifts we use the simplest
parametrizations, which are suitable in the physical region of the
reaction $\gamma\gamma $\,$\to$\,$\pi^0\eta$:
\begin{equation}
e^{i\delta^{bg}_{ab}(s)}=[(1+iF_{ab}(s))/(1-iF_{ab}(s))]^{1/2},
\end{equation} where
\begin{eqnarray}
F_{\pi\eta}(s)=\frac{\sqrt{1-m^{(+)\,2}_{\pi\eta}/s}\left(c_0+c_1
\left(s-m^{(+)\,2}_{\pi\eta}\right)\right)}
{1+c_2\left(s-m^{(+)\,2}_{\pi\eta}\right)^2}, && \\
F_{K\bar K}(s)=f_{K\bar K}\sqrt{s}
\left(\rho_{K^+K^-}(s)+\rho_{K^0\bar K^0}(s)\right)/2, &&
\\ F_{\pi\eta'}(s)=f_{\pi\eta'}\sqrt{s-m^{(+)\,2}_{\pi\eta'}}.
\mbox{\qquad\qquad} && \end{eqnarray}

We now turn to the curves shown in Fig. \ref{ABCDEF}(b) and discuss
the contribution from the heavy (probably existent \cite{PDG08})
$a'_0$ resonance with a mass $m_{a'_0}\approx$\,1.4 GeV (see variant
1 in Table \ref{Tab1}). There is a distinct enhancement in the
region of 1.4 GeV in the cross section corresponding to the
amplitude $M^{\mbox{\scriptsize{direct }}}_ {\mbox{\scriptsize{res}}
}(s)$ (see the short-dashed line). This enhancement turns into the
shoulder in the cross section that corresponds to the total
resonance contribution (see the dotted line), i.e., the combined
contribution from the amplitude $M^{\mbox{\scriptsize{direct }}}_
{\mbox{\scriptsize{res}}}(s)$ and the rescattering amplitudes
proportional to $T^{res}_{ab\to\pi^0\eta}(s)$; see Eq. (\ref{Tres}).
The proper resonance signal is practically absent in the region of
1.4 GeV in the total cross section $\sigma_0$ (see solid line 2)
involving additionally the $\gamma\gamma$\,$\to$\,$\pi^0\eta$ Born
contribution and the rescattering $\gamma\gamma$\,$\to$\,$\pi^0\eta
$\,$\to$\,$\pi^0\eta$ caused by the background elastic amplitude
$\pi^0\eta$\,$\to$\,$ \pi^0\eta$. Thus, there is strong destructive
interference between the different contributions, which hides the
$a'_0$ resonance in the $\gamma\gamma$\,$\to$\,$\pi^0\eta$ cross
section \cite{FN5}. Nevertheless, in many respects owing to the
$a'_0$, we succeed in modeling a large smooth background required by
the Belle data \cite{Ue09} both under the $a_2(1320)$ and between
$a_0(980)$ and $a_2(1320)$ resonances.

Note that the $a'_0$ mass is not well determined for the existing
compensations and the correlations between the fitted parameters.
Even if one puts $m_{a'_0}\approx $\,1.28 GeV (instead of 1.4 GeV),
then one can obtain a reasonable description of the data. The fit
variant (denoted as variant 2) demonstrates this fact by means of
Figs. \ref{ABCDEF}(d)--\ref{ABCDEF}(f), \ref{TEPS}(c), and
\ref{TEPS}(d) and Table \ref{Tab1}. Moreover, this variant gives
some idea of model uncertainties of the other parameter values and
possible variations of the pattern of the $\pi\eta$ scattering
amplitude \cite{FNN2}.

It is interesting to consider the $\gamma\gamma$\,$\to$\,$\pi^0
\eta$ cross section attributed only to the resonance contributions
and, following \cite{AS88,AS05,AS08,AS09}, to determine the width of
the $a_0(980)$\,$ \to$\,$\gamma\gamma$ decay averaged over the
resonance mass distribution in the $\pi\eta$ channel:
\begin{equation}\langle\Gamma_{a_0\to\gamma\gamma}\rangle_{\pi\eta}=
\int\limits_{0.9\mbox{\,\scriptsize{GeV}}}^{1.1\mbox{\,\scriptsize{GeV}}}
\frac{s}{4\pi^2}\sigma_{\mbox{\scriptsize{res}}}(\gamma\gamma\to
\pi^0\eta;s)d\sqrt{s}\end{equation} [the integral is taken over the
region occupied by the $a_0(980)$ resonance]. This quantity is an
adequate characteristic of the coupling of the $a_0(980)$ resonance
with a $\gamma\gamma$ pair. Taking into account in
$\sigma_{\mbox{\scriptsize{res}}}$ the contributions from all the
rescatterings and direct transitions into $\gamma\gamma$, we obtain
$\langle\Gamma_{a_0\to(K\bar K+\pi\eta+\pi\eta'+\mbox
{\scriptsize{direct}})\to\gamma\gamma}\rangle_{\pi\eta}$\,$\approx$\,0.4
keV. For the rescatterings only $\langle\Gamma_{a_0\to(K\bar
K+\pi\eta+\pi\eta')\to\gamma\gamma}
\rangle_{\pi\eta}$\,$\approx$\,0.23 keV and for the direct
transitions only $\langle\Gamma^{\mbox{\scriptsize{direct}}}_{a_0\to
\gamma\gamma} \rangle_{\pi\eta}$\,$\approx$\,0.028 keV. These
estimates correspond to variant 1. Variant 2 gives the close values.

\section{CONCLUSION}

We have analyzed in detail the Belle data on the cross section for
the reaction $\gamma\gamma $\,$\to$\,$\pi^0\eta$ with the use of a
sufficiently full dynamical model and have shown that the
experimentally observed pattern is the result of the combination of
many dynamical factors. In particular, the performed analysis allows
the definite conclusion concerning the dominance of the rescattering
mechanisms $a_0(980)$\,$\to$\,$(K\bar
K+\pi^0\eta+\pi^0\eta')$\,$\to$\,$\gamma\gamma$, i.e., four-quark
transitions, in the decay $a_0(980)$\,$\to$\,$\gamma\gamma$. In
turn, this gives a new argument in favor of the $q^2\bar q^2$ nature
of the $a_0(980)$ resonance. As to the ideal $q\bar q$ model
prediction for the two-photon decay widths of the $f_0(980)$ and
$a_0(980)$ mesons, $\Gamma_{ f_0\to\gamma\gamma
}/\Gamma_{a_0\to\gamma\gamma}=25/9$, it is excluded by experiment.
One more result of our analysis consists in the preliminary
information obtained for the first time on the $S$ wave amplitude of
the reaction $\pi^0\eta$\,$\to$\,$\pi^0\eta$.

The investigations of the mechanisms of the reactions
$\gamma\gamma$\,$\to$\,$\pi^+\pi^-$, $\gamma\gamma$\,$\to$\,$\pi^0
\pi^0$, $\gamma\gamma$\,$\to$\,$\pi^0 \eta$, $\gamma\gamma$\,$\to
$\,$K^+K^-$, and $\gamma\gamma$\,$\to$\,$K^0 \bar K^0$ are
undoubtedly an important constituent of physics of light scalar
mesons. The Belle Collaboration has investigated the reactions
$\gamma\gamma$\,$\to$\,$\pi^+\pi^-$, $\gamma\gamma$\,$\to$\,$
\pi^0\pi^0$, and $\gamma\gamma$\,$\to$\,$\pi^0\eta$ with the highest
statistics. However, similar information is still lacking for the
processes $\gamma\gamma$\,$\to$\,$K^+K^-$ and $\gamma\gamma$\,$\to
$\,$K^0\bar K^0$. The $S$ wave contributions near thresholds of
these two channels are not clearly understood \cite{AS09ar,AS92,
AS94}.

To conclude, we point to the promising possibility of investigating
the nature of the light scalar mesons $\sigma(600)$, $f_0(980)$, and
$a_0(980)$ in the $\gamma\gamma^*$ collisions, where $\gamma^*$ is a
virtual photon with virtuality $Q^2$. If they are four-quark states,
their contributions to the $\gamma\gamma^*$\,$\to$\,$\pi^+\pi^-$,
$\gamma\gamma^*$\,$\to$\,$\pi^0\pi^0$, and $\gamma\gamma^*$\,$\to
$\,$\pi^0\eta$ cross sections should decrease with increasing $Q^2$
more rapidly than the contributions from the classical tensor mesons
$f_2(1270)$ and $a_2(1320)$. A similar behavior of the contribution
from the $q^2\bar q^2$ exotic resonance state with
$I^G(J^{PC})$\,=\,$2^+(2^{++})$ \cite{ADS82} to the
$\gamma\gamma^*$\,$\to$\,$\rho^0\rho^0$ and $\gamma\gamma^*$\,$\to
$\,$\rho^+\rho^-$ cross sections was recently observed by the L3
Collaboration \cite{L3}.
$$\mbox{\small \bf ACKNOWLEDGMENTS}$$

This work was supported in part by the RFFI Grant No. 10-02-00016
from the Russian Foundation for Basic Research.


\begin{thebibliography}{99}
\bibitem{FN1}    Such an interpritation of the $\gamma\gamma$\,$
\to$\,$\pi^0\eta$ data testifies in favor of the $q^2\bar q^2 $
nature of the $a_0(980)$ state and agrees with the previous
interpretation \cite{AS05,AS08,AS09ar} of the high-statistics Belle
data on the reactions $\gamma\gamma$\,$\to$\,$\pi^+\pi^-$
\cite{Mo07} and $\gamma\gamma$\,$\to$\,$ \pi^0\pi^0$ \cite{Ue08} and
with the properties of the $\sigma(600)$ and $f_0(980)$ resonances
which are the partners  of the $a_0(980)$. In particular, in Refs.
\cite{AS05,AS08} it was shown that the $\sigma(600)$\,$
\to$\,$\gamma \gamma$ and $f_0(980)$\,$\to$\,$\gamma \gamma$ decays
are due basically to the rescattering mechanisms, i.e., to
four-quark transitions such as $\sigma(600)$\,$\to$\,$
\pi^+\pi^-$\,$\to$\,$\gamma\gamma$ and $f_0(980)$\,$\to$\,$
K^+K^-$\,$\to$\,$\gamma\gamma$. As for the $\sigma(600)$ and
$f_0(980)$ direct coupling constants to $\gamma\gamma$, they are
small. A similar conclusion about the direct coupling of the
$\sigma(600)$ to $\gamma\gamma$ has been also obtained in Ref.
\cite{MNO}. Thus, the mechanisms of the $\gamma\gamma$ decays of
light scalars $\sigma(600)$, $f_0(980)$, and $a_0(980)$ differ in
principle from the mechanism of the $\gamma\gamma$ decays of
classical $P$ wave tensor $q\bar q$ mesons $f_2(1270)$, $a_2(1320)$,
and $f'_2(1525)$, which is the direct annihilation transition $q\bar
q$\,$\to$\,$\gamma\gamma$ \cite{Ac08}.
\bibitem{FNN1}   Properly calculate the accuracy of this approximation is rather
difficult. From the chiral Lagrangian standpoint, the off-shell
effects are related with the higher order corrections. The
phenomenological form factors introduced into the Born exchange
amplitudes effectively take into account the deviations caused by
such corrections.
\bibitem{FN2}    For the elementary one-kaon exchange
$\widetilde{I}^{K^+}_{K^+K^-}(s)=8\alpha(\frac{m^2_{K^+}}{s}[\pi+i
\ln\frac{1+\rho_{K^+K^-}(s)}{1-\rho_{K^+K^-}(s)}]^2-1)$ for
$s\geq4m^2_{K^+}$ and $\widetilde{I}^{K^+}_{K^+K^-}(s)=8\alpha(
\frac{m^2_{K^+}}{s}[\pi-2\arctan|\rho_{K^+K^-}(s)|]^2-1)$ for $0\leq
s\leq4m^2_{K^+}$ \cite{AS88}; $\alpha$\,=\,1/137. In this work, to
calculate the function $\widetilde{I}^{K^+}_{K^+K^-}(s)$ we use the
amplitude $M^{\mbox{\scriptsize{Born}}\,K^+}_{00} (s)=\frac{16
\pi\alpha
m^2_{K^+}}{s\rho_{K^+K^-}(s)}\left[\ln\frac{1+\rho_{K^+K^-}(s)}
{1-\rho_{K^+K^-}(s)}-\ln\frac{1+\rho_{K^+K^-}
(s;x_2)}{1-\rho_{K^+K^-}(s;x_2) }\right]$, where the second term in
square brackets appears for the form factor $G_{K^+}(t,u)=
\frac{1}{s}\left[\frac{m^2_{K^+}-t}{1-(u-m^2_{K^+})/x^2_2}+
\frac{m^2_{K^+}-u}{1-(t-m^2_{K^+})/x^2_2}\right]$, which suppresses
the elementary one-kaon exchange amplitudes with increasing
$\sqrt{s}$;
$\rho_{K^+K^-}(s;x_2)$\,=\,$\rho_{K^+K^-}(s)/(1+2x^2_2/s)$ and $x_2$
is an input parameter. The form factor of such a type was proposed
in the work \cite{Po86}. Note that the form factor is introduced
through replacing the elementary one-kaon exchange amplitudes
$\widehat{M}_0^{\mbox{\scriptsize{Born}}\,K^+}(s,\theta)$\,=\,$(32
\pi\alpha m^2_{K^+}/s)/[1-\rho^2_{K^+K^-}(s)\cos^2\theta]$,
$\widehat{M}_2^{\mbox{\scriptsize{Born}}\,K^+}(s,\theta)$\,=\,$8\pi
\alpha\rho^2_{K^+K^-}(s)\sin^2\theta/[1-\rho^2_{K^+K^-}(s)$
$\times\cos^2 \theta]$ by
$M^{\mbox{\scriptsize{Born}}\,K^+}_\lambda(s,\theta)=G_{K^+}(t,u)
\widehat{M}_\lambda^{\mbox{\scriptsize{Born}}\,K^+}(s,\theta)$,
which keeps gauge invariance beyond the tree approximation
\cite{Po86}. The necessity of the inclusion of the form factors in
the Born amplitudes of the elementary one-pion and one-kaon
exchanges in the reactions $\gamma\gamma\to\pi^+\pi^-$ and
$\gamma\gamma\to K^+K^-$ was recently discussed in Refs.
\cite{AS08,AS09}, see also Refs. \cite{AS92,Po86,Jo86,MPBB,AS94}. In
particular, the estimate $x_2\approx1.75$ GeV was obtained in Ref.
\cite{AS08} from the simultanious analysis of the Belle data on the
$\gamma\gamma$\,$\to$\,$\pi^+\pi^-$ and $\gamma\gamma\to\pi^0\pi^0$
cross sections; this estimate is used in this work.
\bibitem{FN4}   The negative background phase is not news. It was
shown in Ref. \cite{AS94a} that in the linear $\sigma$ model there
is a negative background phase in the $S$ wave $\pi\pi$ scattering
amplitude with $I$\,=\,0 which hides the $\sigma$ meson with the
result that the $\pi\pi$ scattering phase shift $\delta^0_0$ does
not pass over $90^0$ at the putative resonance mass. It has been
made clear that shielding wide lightest scalar mesons in chiral
dynamics is very natural \cite{Ac08,AK06,AS05, AS08}.
\bibitem{FN5}   A similar compensation between the background and $a'_0$
resonance contributions has been also obtained in the
phenomenological fit performed by the authors of the Belle
experiment \cite{Ue09}.
\bibitem{FNN2}  It should be emphasized that the predictions for the elastic
$\pi\eta$ scattering amplitude $T^1_0(s)$ are very preliminary for
the large theoretical uncertainties associated with the vector
exchanges. The strong model dependence of $T^1_0(s)$ is adequately
illustrated in Fig. \ref{TEPS}.
\end{thebibliography}
\end{document}